\documentclass[a4paper]{jpconf}
\usepackage{graphicx}
\begin{document}
\title{Dissipative hydrodynamic effects on the quark-gluon plasma at finite baryon density}

\author{Akihiko Monnai$^{1,2}$}

\address{${}^1$ Department of Physics, The University of Tokyo, Tokyo 113-0033, Japan}
\address{${}^2$ Theoretical Research Division, Nishina Center, RIKEN, Wako 351-0198, Japan}

\ead{monnai@nt.phys.s.u-tokyo.ac.jp}

\begin{abstract}
The quark-gluon plasma behaves as a relativistic viscous fluid in high-energy heavy ion collisions. I develop a causal dissipative hydrodynamic model at finite baryon density for RHIC and LHC to estimate the net baryon rapidity distribution. The net baryon number is found to be carried to forward rapidity by the flow, effectively enhancing the transparency of the collisions. This suggests that the energy available for the production of a hot medium could be larger than that naively implied from experimental data. Also the distribution can be sensitive to baryon dissipation as much as to shear and bulk viscosity.
\end{abstract}

\section{Introduction}
The quark-gluon plasma (QGP) \cite{Yagi:2005yb} is a many-body system of quarks and gluons deconfined from hadrons at high energies. It is considered to be experimentally producible in high-energy heavy ion collisions at the Relativistic Heavy Ion Collider (RHIC) at the Brookhaven National Laboratory and the Large Hadron Collider (LHC) at the European Organization for Nuclear Research, providing great opportunities and challenges for developing realistic theories of the hot and dense QCD matter. One of the most distinctive characteristics of the QGP implied from the experimental data is that it behaves as a relativistic fluid with extremely small viscosity. This allows one to construct a hydrodynamic effective model of the high-energy nuclear collisions. Modern hydrodynamic studies incorporate viscosity and fluctuation for precision physics \cite{Schenke:2010rr}.

The net baryon number, so far, has been neglected in viscous hydrodynamic studies with an evolving flow. This could be due to the complexity of dissipative processes in causal relativistic formalisms and the boost invariant ansatz that fails to take account of the net baryon number conserved at forward rapidity. It should be noted that finite-density effects are generally more prominent near the beam rapidity since the net baryon number originates from the remnant of the shattered nuclei. Introducing the net baryon to the model would enable one to understand the matter-to-antimatter ratios of the particle spectra as well as the transport properties of finite-density media. 
The analyses of the finite-density matter via dissipative hydrodynamics would also be of great help in exploring the QCD phase diagram for the search of the critical point, given that first principle approaches have difficulties for calculating systems with a finite baryon chemical potential. As the beam energy scans are being performed and planned at RHIC, GSI, NICA and J-PARC, it is important to develop a finite-density model to analyze the data and to examine the validity of the model for those experiments.

The net baryon distribution is quantified with \textit{baryon stopping} defined by the mean rapidity loss $\langle \delta y \rangle = y_p - \langle y \rangle$ \cite{Busza:1983rj,Videbaek:1995mf}. Here $y_p$ is the rapidity of the incoming projectile and 
\begin{equation}
\langle y \rangle = \int _0^{y_p} y \frac{dN_{B-\bar{B}}(y)}{dy} dy  
\bigg/ \int _0^{y_p} \frac{dN_{B-\bar{B}}(y)}{dy} dy,
\end{equation}
is the average rapidity of the net baryon density. When $\langle \delta y \rangle = 0$, the collision is completely transparent and no hot medium would be produced. On the other hand, $\langle \delta y \rangle = y_p$ suggests that the nuclei comes to a complete halt at the collision, losing all the kinetic energy for the medium. Thus this quantify is related to the magnitude of the energy loss of the colliding nuclei that is available for the QGP production. Here I would like to focus on dissipative hydrodynamic effects on the net baryon distribution and the mean rapidity loss \cite{Monnai:2012jc}.

\section{Dissipative hydrodynamic model}
The energy momentum tensor $T^{\mu \nu}$ and the net baryon current $N^\mu_B$ are the conserved quantities in the system of interest. Hydrodynamics introduces macroscopic flow $u^\mu$ which is defined here as the local energy flux. Then the tensor decomposition in terms of the flow yields
\begin{eqnarray}
T^{\mu \nu} = (e_0 + P_0 + \Pi )  u^\mu u^\nu - (P_0 + \Pi) g^{\mu \nu} + \pi^{\mu \nu}, \ N^\mu_B = n_{B0} u^\mu + V^\mu ,
\end{eqnarray}
where $e_0$, $P_0$, $n_{B0}$, $\Pi$, $\pi^{\mu \nu}$ and $V^\mu$ are the energy density, the hydrostatic pressure, the net baryon density, the bulk pressure, the shear stress tensor and the baryon dissipation current, respectively. The system is described by energy-momentum conservation $\partial_\mu T^{\mu \nu} = 0$, baryon number conservation $\partial_\mu N_B^\mu = 0$ and the law of increasing entropy $\partial_\mu s^\mu \geq 0$. By introducing second order corrections to the entropy current $s^\mu$ in terms of the dissipative quantities, the semi-positive definite condition yields causal hydrodynamic equations of motion \cite{Israel:1979wp,Monnai:2010qp}:

\begin{eqnarray}
\Pi &=& -\zeta \nabla _\mu u^\mu - \zeta_{\Pi \delta e} D\frac{1}{T} + \zeta_{\Pi \delta n_B} D\frac{\mu_B}{T} - \tau_\Pi D \Pi \nonumber \\
&+& \chi_{\Pi \Pi}^{a} \Pi D\frac{\mu _B}{T} + \chi_{\Pi \Pi}^b \Pi D \frac{1}{T} + \chi_{\Pi \Pi}^c \Pi \nabla _\mu u^\mu + \chi_{\Pi V}^{a} V_\mu \nabla ^\mu \frac{\mu_B}{T} + \chi_{\Pi V}^b V_\mu \nabla ^\mu \frac{1}{T} \nonumber \\
&+& \chi_{\Pi V}^c V_\mu D u ^\mu + \chi_{\Pi V}^d \nabla ^\mu V_\mu + \chi_{\Pi \pi} \pi _{\mu \nu} \nabla ^{\langle \mu} u^{\nu \rangle} , 
\\
V^\mu &=& \kappa_{V} \nabla ^\mu \frac{\mu _B}{T} - \kappa _{V W} \bigg( \nabla ^\mu \frac{1}{T} + \frac{1}{T} D u^\mu \bigg) - \tau _{V} \Delta^{\mu \nu} D V_\nu \nonumber \\
&+& \chi_{V V}^{a} V^\mu D \frac{\mu _B}{T} + \chi_{V V}^b V^\mu D \frac{1}{T} + \chi_{V V}^c V^\mu \nabla _\nu u^\nu + \chi_{V V}^d V^\nu \nabla _\nu u^\mu + \chi_{V V}^e V^\nu \nabla ^\mu u_\nu \nonumber \\
&+& \chi_{V \pi}^{a} \pi^{\mu \nu} \nabla_\nu \frac{\mu _B}{T} + \chi_{V \pi}^b \pi^{\mu \nu} \nabla_\nu \frac{1}{T} + \chi_{V \pi}^c \pi ^{\mu \nu} D u_\nu + \chi_{V \pi}^d \Delta^{\mu \nu} \nabla ^\rho \pi _{\nu \rho} \nonumber \\
&+& \chi_{V \Pi}^{a} \Pi \nabla ^\mu \frac{\mu _B}{T} + \chi_{V \Pi}^b \Pi \nabla ^\mu \frac{1}{T} + \chi_{V \Pi}^c \Pi D u^\mu + \chi_{V \Pi}^d \nabla ^\mu \Pi , \label{V}\\
\pi^{\mu \nu} &=& 2 \eta \nabla ^{\langle \mu} u^{\nu \rangle} - \tau_\pi D \pi^{\langle \mu \nu \rangle}  \nonumber \\
&+& \chi _{\pi \Pi} \Pi \nabla ^{\langle \mu} u^{\nu \rangle} + \chi_{\pi \pi}^{a} \pi^{\mu \nu} D \frac{\mu_B}{T} + \chi_{\pi \pi}^b \pi^{\mu \nu} D \frac{1}{T} + \chi_{\pi \pi}^c \pi ^{\mu \nu} \nabla _\rho u^\rho + \chi_{\pi \pi}^d \pi ^{\rho \langle \mu} \nabla _\rho u^{\nu \rangle} \nonumber \\
&+& \chi_{\pi V}^{a} V^{\langle \mu} \nabla ^{\nu \rangle} \frac{\mu_B}{T} + \chi_{\pi V}^b V^{\langle \mu} \nabla ^{\nu \rangle} \frac{1}{T} + \chi_{\pi V}^c V^{\langle \mu} D u^{\nu \rangle} + \chi_{\pi V}^d \nabla ^{\langle \mu} V ^{\nu \rangle} .
\end{eqnarray}
Here $D = u^\mu \partial_\mu$ and $\nabla_\mu = \partial_\mu - u_\mu D$. The angular bracket on the indices denotes traceless symmetrization. $T$ and $\mu_B$ are the temperature and the chemical potential, respectively. $\zeta$, $\kappa_V$ and $\eta$ are the bulk viscosity, the baryon charge conductivity and the shear viscosity, which characterize scalar, vector and tensor dissipative processes in a medium at the linear order. $\zeta_{\Pi \delta e}$ and $\zeta_{\Pi \delta n_B}$ are the bulk-energy and the bulk-baryon cross viscosities and $\kappa_{VW}$ is the baryon-heat cross conductivity, which should satisfy the Onsager reciprocal relations. It is note-worthy that they can be negative. $\tau_\Pi$, $\tau_V$ and $\tau_\pi$ are the relaxation times, which are required for causality and stability in relativistic systems. $\chi$ are second order transport coefficients. 

One has to introduce the equation of state, the transport coefficients and the initial conditions to construct a model for the high-energy heavy ion collisions. The equation of state is obtained via Taylor expansion with the results from latest lattice QCD calculations \cite{Borsanyi:2010cj,Borsanyi:2011sw}. The transport coefficients are obtained from the AdS/CFT approaches \cite{Kovtun:2004de,Natsuume:2007ty}, the non-equilibrium statistical operator method \cite{Hosoya:1983id} and phenomenological analyses based on the dimensionality, the matter-antimatter symmetry and the semi-positive definiteness of the transport coefficient matrices \cite{Monnai:2012jc}. It should be noted that the coefficients{ are introduced to investigate qualitative responses of the system. The initial conditions at $\tau_0$ = 1 fm are based on the color glass theory; the energy density is from the Nara Monte-Carlo version \cite{Drescher:2006ca,Drescher:2007ax} of Kharzeev-Levin-Nardi model \cite{Kharzeev:2002ei,Kharzeev:2004if} and the net baryon density from the distribution of the valence quarks \cite{MehtarTani:2008qg,MehtarTani:2009dv}. The list of individual model input can be found in Table~\ref{tc}, where $\chi^{(2)}_B$ is the quadratic baryon fluctuation, $s$ is the entropy density and $c_s = [(\partial e_0 / \partial P_0 )_{s/n_{B0}}]^{1/2}$ is the sound velocity:
\begin{table}[h]
\small
\caption{\label{tc}Input to dissipative hydrodynamic model} 
\begin{center} 
\begin{tabular}{ll} 
\br 
Equation of state&Description\\ 
\mr
EoS at finite $\mu_B$& $P_0(e_0, n_{B0}) = P_0 (e_0, 0) + (\mu_B^2 T^2/2)\chi^{(2)}_B (e_0,0)$\\  
\br 
Transport coefficients&Description\\ 
\mr 
Shear viscosity &$\eta = s/4\pi$\\ 
Effective bulk viscosity$^{*}$ &$\zeta_\mathrm{eff} = (5/2) [(1/3) - c_s^2] \eta$\\ 
Baryon charge conductivity&$\kappa_V = c_V (\partial \mu_B / \partial n_B)^{-1}_T / 2\pi$\\ 
Baryon-heat conductivity&$\kappa_{VW} = c_{VW} [n_{B0}/(e_0+P_0)] (5\eta T \kappa_V)^{1/2}$\\ 
\br
Initial conditions&Description\\ 
\mr
Energy density &$e_0(\tau_0,\eta_s) = (1/ S_\mathrm{area})dE_T/\tau_0 d\eta_s$\\  
Net baryon density &$n_{B0}(\tau_0,\eta_s) = (1/S_\mathrm{area})dN_{B-\bar{B}}/\tau_0 d\eta_s$\\  
\br 
\end{tabular} 
\end{center}
*The contributions of the scalar cross coefficients are merged into the effective bulk viscosity.
\end{table}

In the present analyses, I discuss dissipative hydrodynamic expansion in the longitudinal direction by integrating out the transverse geometry for the numerical analyses. It is note-worthy that the dependence of the net baryon yields on transverse geometry is shown to be very small experimentally in the analyses of centrality dependence \cite{Adler:2003cb}.

\section{Results}
Fig.~\ref{fig:1} shows the net baryon rapidity distributions for the Au-Au collisions at RHIC and the Pb-Pb collisions at LHC for the most central 0-5 \% events. The parameters for the conductivities are $c_V = 1$ and $c_{VW} = 0$ for the moment. The contribution of hadron resonance up to 2 GeV is considered at the freeze-out temperature $T_f= 0.16$ GeV. The experimental data points are scaled results from the corresponding net proton distribution by the BRAHMS experiments \cite{Bearden:2003hx}.
One can see that the net baryon is carried to the forward rapidity region in hydrodynamic evolution. This is mainly due to the pressure gradient, to which the net baryon density is coupled via the flow. The effect is slightly moderate for the shear and bulk viscous case because the longitudinal pressure is effectively reduced by the viscosities. Baryon dissipation further steepens the distribution as it is induced by the chemical gradient into the mid-rapidity region. For the current parameter settings, the hydrodynamic results for RHIC roughly agrees with the experimental data. The off-equilibrium corrections are more visible at RHIC than at LHC. 
\begin{figure}
\includegraphics[width=18pc]{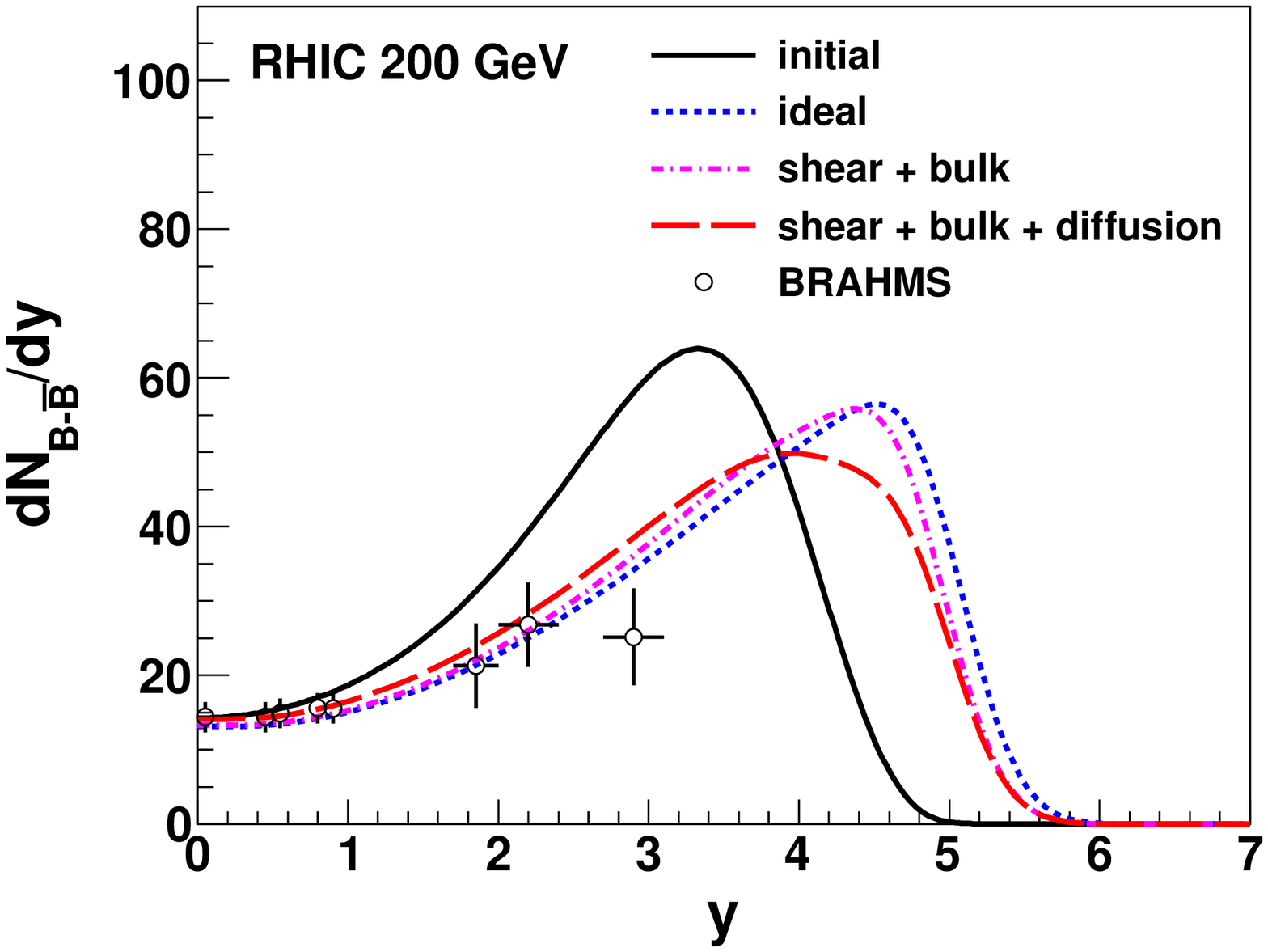}
\includegraphics[width=18pc]{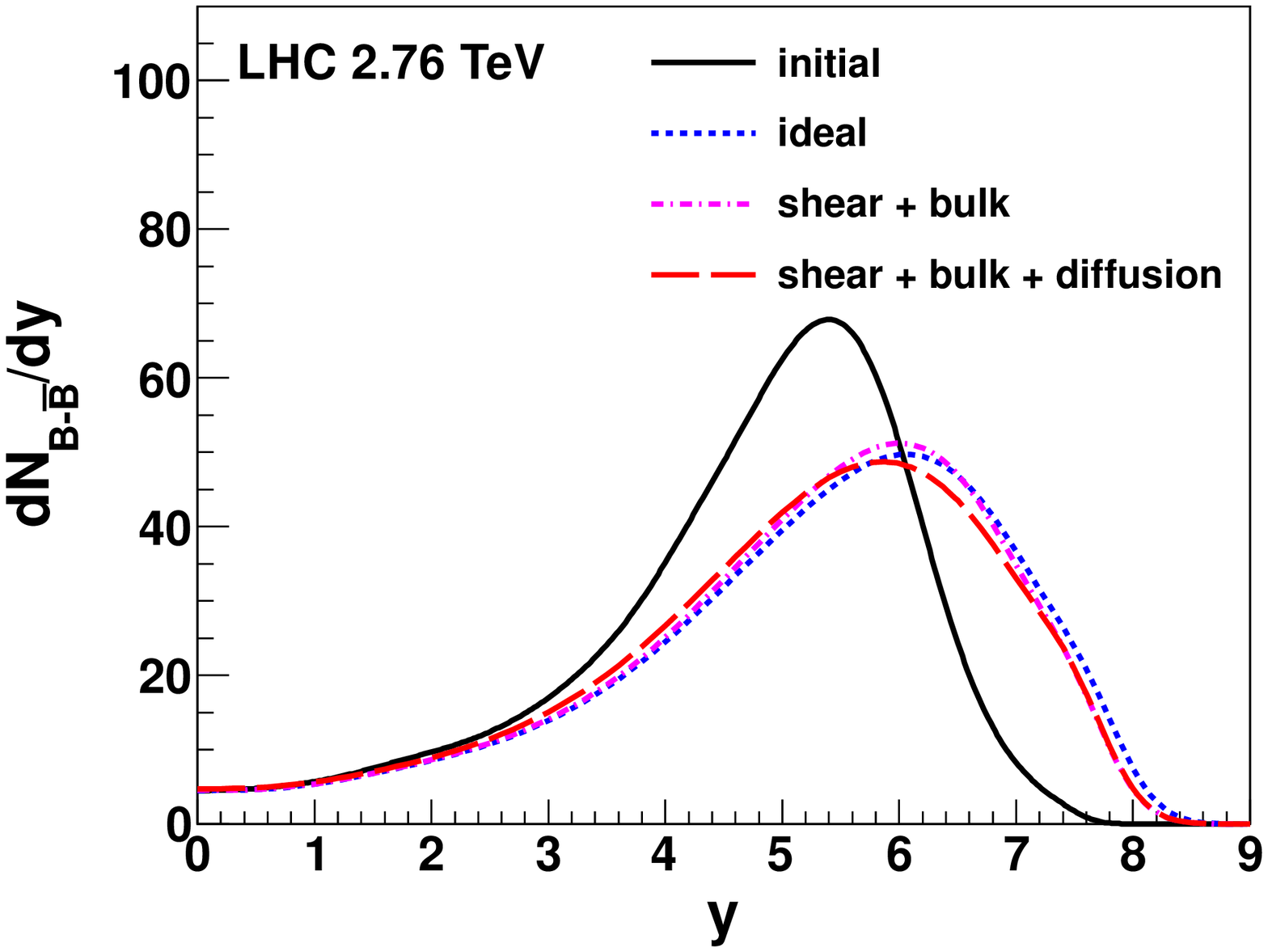}
\caption{\label{fig:1} \small The net baryon distributions of the initial condition and ideal, viscous and dissipative hydrodynamic results at RHIC (left) and at LHC (right) are shown as solid, dotted, dash-dotted and dashed curves, respectively. The experimental data points are from the BRAHMS Collaboration \cite{Bearden:2003hx}.}
\end{figure}

The hydrodynamic evolution reduces the mean rapidity loss from $\langle \delta y \rangle_\mathrm{initial} = 2.67$ to $\langle \delta y \rangle_\mathrm{hydro} = 2.09, 2.16$ and 2.26 for the ideal, the viscous and the dissipative systems at RHIC. Here the dissipative results include effects of baryon diffusion in addition to those of shear and bulk viscosities. It shows that the transparency of the collision is sizably enhanced as a medium effect. This could explain the abrupt deviation of the RHIC data from the simple extrapolation of the AGS and SPS results (Fig.~\ref{fig:2}) since the hot medium is created at the heavy ion collisions with sufficiently high energies. The kinetic energy losses after the hydrodynamic evolutions are also about 10-15\% less than those of the initial conditions, suggesting that more energy is available for the QGP production at the time of collision than naively implied from the experimental data. 
At LHC, when $\langle \delta y \rangle_\mathrm{initial} = 3.36$, the rapidity loss is reduced to $\langle \delta y \rangle_\mathrm{hydro} = 2.92$ for the dissipative hydrodynamic case. It should be emphasized that the result is dependent on the choice of the initial conditions and the transport coefficients. 

\begin{figure}
\begin{minipage}{18pc}
\includegraphics[width=18pc]{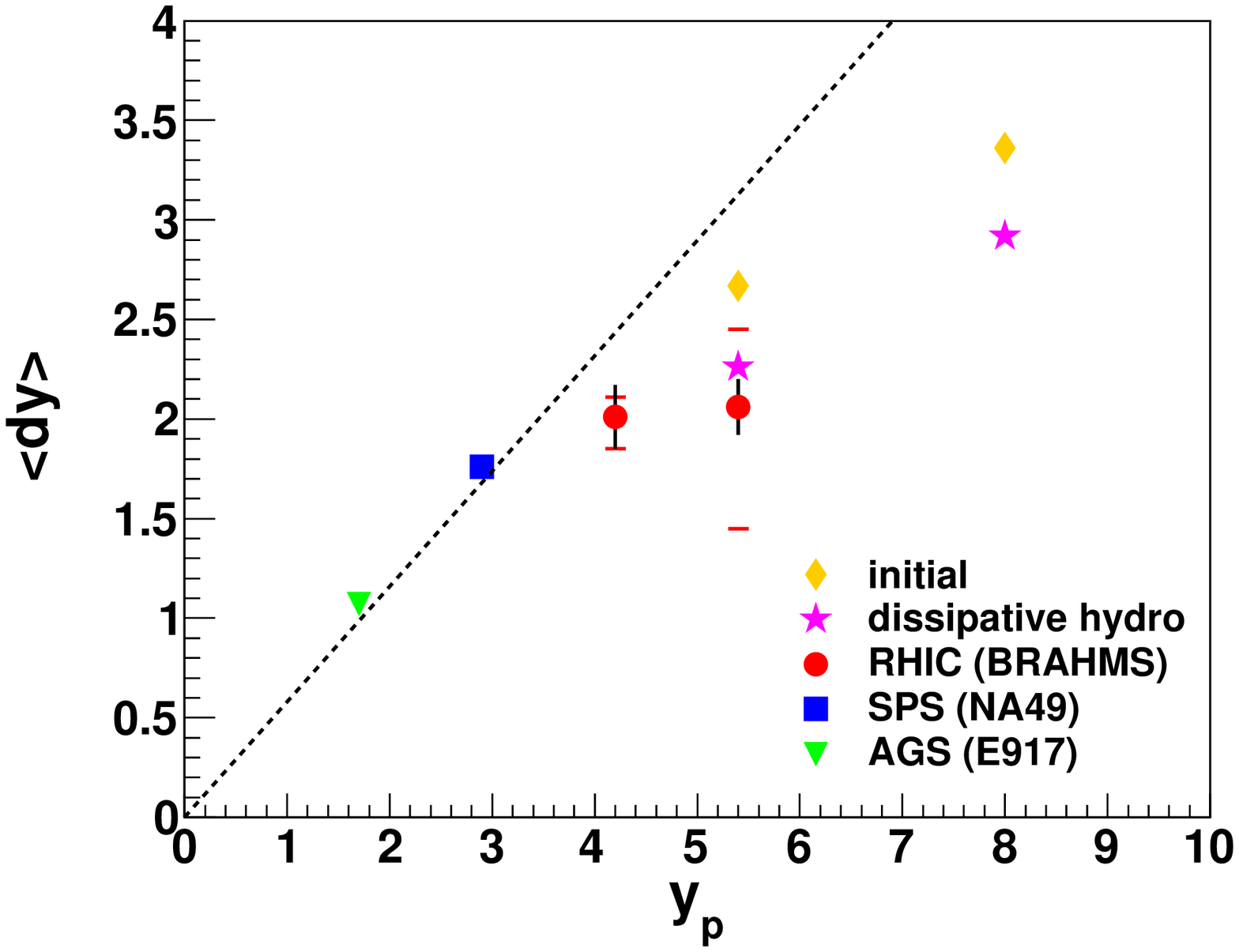}
\caption{\label{fig:2} \small Mean rapidity losses at AGS \cite{Back:2000ru}, SPS \cite{Appelshauser:1998yb} and RHIC \cite{Bearden:2003hx,Arsene:2009aa} experiments and those in dissipative hydrodynamic calculations for RHIC and LHC. 
The dotted line is a linear extrapolation of the low energy results \cite{Bearden:2003hx}.}
\end{minipage}\hspace{2pc}%
\begin{minipage}{18pc}
\includegraphics[width=18pc]{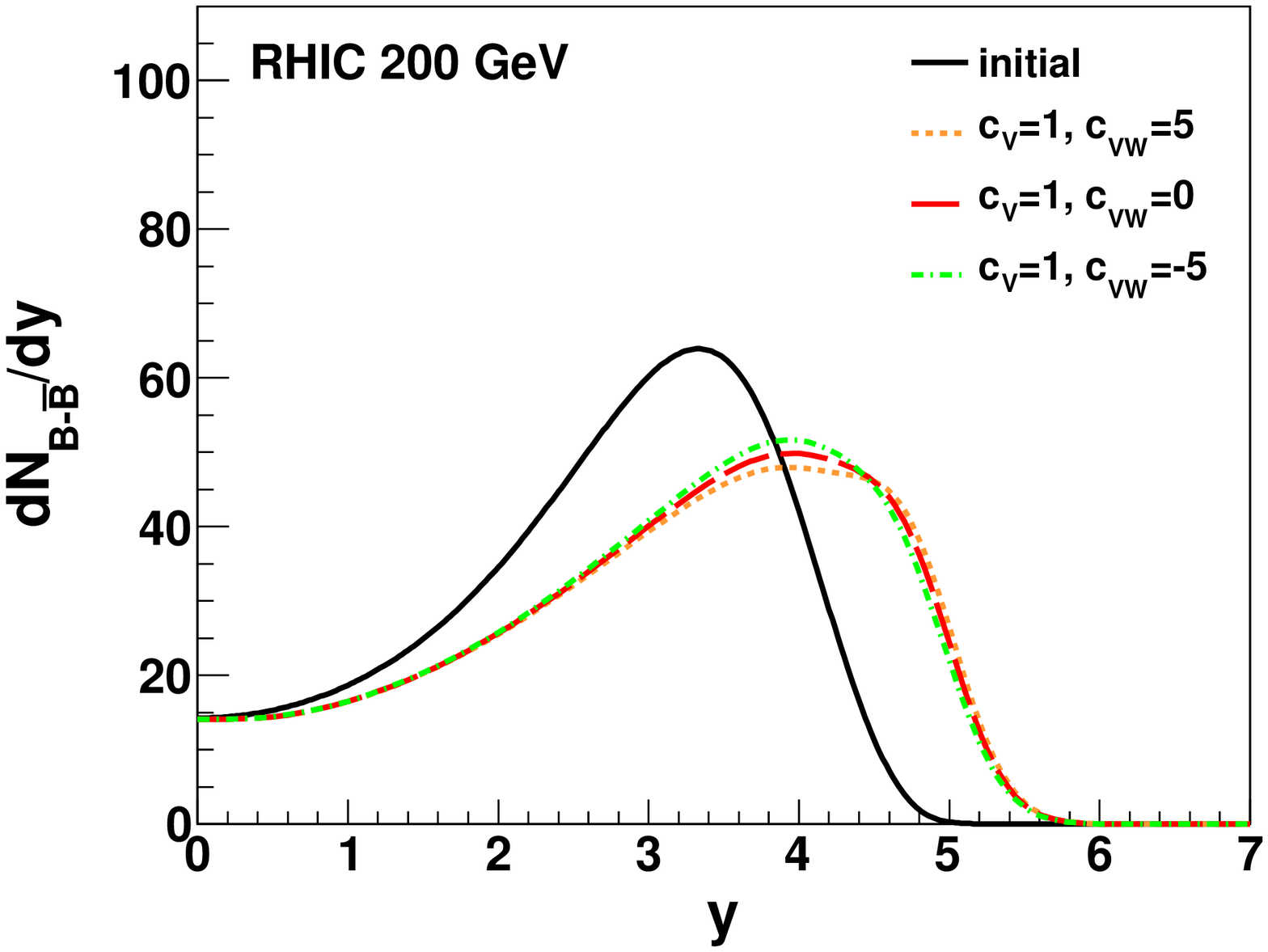}
\caption{\label{fig:3} \small The net baryon distribution of the initial condition and dissipative hydrodynamic results at RHIC with the cross coefficient $c_{VW} = 5, 0$ and $-5$ are shown in solid, dotted, dashed and dash-dotted curves, respectively.}
\end{minipage} 
\end{figure}

Finally, the net baryon distributions in the presence of the baryon-heat cross-coupling term is shown in Fig.~\ref{fig:3}. The parameter for the cross coefficient is chosen as $c_{WV} = 5,0$ and $-5$. The term induces Soret effect, a chemical diffusion process by the thermal gradient. The numerical results indicate that the effect is rather small. This is due to the fact that the cross conductivity has to stay small when the chemical potential is not large as $\kappa_{VW} (- \mu_B) = - \kappa_{VW} (\mu_B)$ follows from the matter-antimatter symmetry condition $V^\mu(- \mu_B) = - V^\mu (\mu_B)$ on Eq.~(\ref{V}). 

\section{Summary}

Non-equilibrium hydrodynamic analyses are extended for the quark-gluon systems with finite baryon density. The finite-density equation of state is constructed in the Taylor expansion method with state-of-art data of lattice QCD calculations and the initial conditions in the color glass-based pictures. The baryon stopping at RHIC and LHC show clear effects of hydrodynamic convection where both viscous and dissipative modifications are qualitatively visible at RHIC. The widening of the net baryon distribution suggests that the kinetic energy loss is effectively reduced during the hydrodynamic evolution. This implies that the energy available for the production of a hot medium is larger initially, and part of the energy is transferred back to the net baryon components at the later stage as the hydrodynamic flow accelerates them. Also the Soret effect would be small in the high-energy nuclear collisions. One would need to include transverse dynamics and realistic transport coefficients for more quantitative discussion.

\ack
The author is grateful for the valuable comments by Hatsuda T. 
The work of A.M. is supported by the Grant-in-Aid for JSPS Research Fellows (No. 10J07647).

\end{document}